\documentclass{ws-procs9x6-cpt16}
\begin{document}

\newcommand{\refeq}[1]{(\ref{#1})}
\def\etal {{\it et al.}}

\def\Dsm{\ensuremath{D_s^-}}
\def\Dsp{\ensuremath{D_s^+}}
\def\DsPdecay{\ensuremath{D_s^+ \rightarrow \phi \pi^{+}}}
\def\DsMdecay{\ensuremath{D_s^- \rightarrow \phi \pi^{-}}}
\def\Bsdecay{\ensuremath{B_s^0 \rightarrow \mu^+ D_s^-}}
\def\Bs{\ensuremath{B^0_s}}
\def\barBs{\ensuremath{\bar{B}^0_s}}

\title{Search for Violation of CPT and Lorentz Invariance in\\
$B^0_s$ Meson Oscillations using the D\O\ Detector}

\author{R.\ Van Kooten}

\address{Department of Physics, Indiana University,
Bloomington, Indiana 47405, USA}

\author{On behalf of the D\O\ Collaboration%
\footnote{http://www-d0.fnal.gov}}

\begin{abstract}
A search is presented for CPT-violating effects in the mixing
of \Bs\ mesons 
using the D0 detector at the Fermilab Tevatron Collider.
The CPT-violating asymmetry in the decay $\Bs \to \mu^\pm D_s^\mp X$  
as a function of sidereal phase is measured.
No evidence for CPT-violating effects is observed and
limits are placed on CPT- and Lorentz-invariance violating 
coupling coefficients.
\end{abstract}

\bodymatter

\section{Introduction}

Lorentz invariance requires that the description of a particle is independent of its direction of motion or boost velocity. The Standard-Model Extension (SME)\cite{CPT2} provides a framework for potential Lorentz and CPT invariance violation (CPTV) suggesting that such violations occur at the Planck scale, but still result in potentially observable effects at currently available collider energies.
In neutral meson systems, the hamiltonian is a $2 \times 2$  matrix relating 
the mass and weak eigenstates.  Mixing between particle and antiparticle
is driven by nonzero off-diagonal matrix elements due to a box 
diagram between $B^0_{(d \thinspace {\mathrm{or}} \thinspace s)}$ and 
$\bar{B}^0_{(d \thinspace {\mathrm{or}} \thinspace s)}$. T (or CP) violation
in mixing can be due to differences between these off-diagonal terms
and results in the two probabilities for oscillation between particle and 
antiparticle not being equal, i.e.,
$P(B^0 \rightarrow \bar{B}^0; t) \neq P(\bar{B}^0 \rightarrow B^0; t)$.
CPT and Lorentz violation involves differences between {\it diagonal} 
terms of this matrix and differences in the probabilities 
$P(B^0 \rightarrow B^0; t) \neq P(\bar{B}^0 \rightarrow \bar{B}^0; t)$ and can be 
expressed with the parameter\cite{cpt_preprint}
\begin{equation}
\xi = \frac{(M_{11} - M_{22}) - \frac{i}{2}(\Gamma_{11} - \Gamma_{22})}{-\Delta m - \frac{i}{2}\Delta \Gamma} \thinspace \approx \frac{\beta^{\mu} \Delta a_{\mu}}{-\Delta m - \frac{i}{2}\Delta \Gamma},
\label{eq:xidef}
\end{equation}
where $\beta^{\mu} = \gamma(1,\vec{\beta})$ is the 4-velocity of the neutral $B$ meson,
$\Delta{m}$ and $\Delta\Gamma$ are the mass and width difference between the
heavy and light mass eigenstates,
and $\Delta a_{\mu} = r_{q_1} a^{q_1}_{\mu} - r_{q_2} a^{q_2}_{\mu}$ with
$r$ being coefficients with $q_1$ and $q_2$ as meson valence quarks and
$a_{\mu}$ being the constant 4-vector in the SME Lagrange density.\cite{SME}
For the $\Bs$-$\barBs$ system, the fractional difference
between the mass eigenvalues is of the order of $10^{-12}$. Due to this,
$\Bs$-$\barBs$ oscillations form an interferometric system that is very
sensitive to small couplings between the valence quarks and a possible
Lorentz-invariance violating field, making it an ideal place to search
for new physics.\cite{CPT1}

\section{Dimuon and \Bs\ semileptonic decay charge asymmetries}

The measurement of the like-sign dimuon asymmetry by the D\O\
Collaboration\cite{dimuon2013} shows  evidence of anomalously large
CP-violating effects.  
This anomalous asymmetry could also
arise from T-invariant CP violation in $\Bs$-$\barBs$ mixing and this 
sensitivity to CPT breaking has been used 
to obtain the first quantitative {\it indirect} measure and limit of CPT violation
in the $\Bs$-$\barBs$ 
system.\cite{vanKooten}

CP- and CPT-violating effects can be explored using the semileptonic decay
$\Bsdecay X$, where \DsMdecay\ and $\phi \to K^+K^-$ (charge
conjugate states are assumed throughout). CP-violating asymmetries
are usually between ``wrong-sign'' decays $\Bs \to \barBs \to
\mu^+\Dsm$, and the D\O\ Collaboration has measured\cite{assl} this 
flavor-specific asymmetry to be
$a^s_{\mathrm{sl}} = [-1.12 \pm 0.74 \thinspace {\mathrm{(stat)}} \pm 0.17 \thinspace {\mathrm{(syst)}}]\%$, i.e., consistent with zero.

\section{D\O\ search for CPT-violating asymmetry}

A D\O\ published analysis\cite{resultpub} explores  the asymmetry between the
 ``right-sign" decays
$\Bs \to \Bs \to \mu^-\Dsp$ and its charge conjugate using 10.4~fb$^{-1}$ of integrated
luminosity collected at the Fermilab Tevatron collider. 
The
CPT-violating parameter is extracted using the asymmetry
\begin{equation}
 \label{raw}
	A = \frac{ N_{+} -  N_{-}}{ N_{+} + N_{-}},
\end{equation}
where  $N_{+}$ [$N_{-}$] is the number of reconstructed $\Bs \to \mu^\pm
D_s^{\mp} X$ events where  $\mathrm{sgn}(\cos\theta)Q>0$
[$\mathrm{sgn}(\cos\theta)Q<0$], $\theta$ is the polar angle between the
\Bs\ reconstructed momentum and the proton beam direction,   
and $Q$ is the charge of the muon. 
The initial state at
production is not flavor tagged, but after experimental
selection requirements, the \Bs\ system is fully mixed, so the
probability of observing a \Bs\ or \barBs\ is essentially equal
regardless of the flavor at production. We assume no CP violation in
mixing,\cite{hfag}
so only about half of the observed \Bs\ have the same flavor as they
had at birth, and observed \Bs\
mesons that have changed their flavor do not contribute to CPTV,
leading to a $\sim 50\%$ dilution in the measured asymmetry.

In the SME, spontaneous Lorentz symmetry breaking generates constant
expectation values for the quark fields that are Lorentz vectors represented by 
$\Delta a_{\mu}$, so any observed CPT violation and the asymmetry above 
should vary in the frame of the
detector with a period of one sidereal day as the direction of the Tevatron's proton 
beam follows the Earth's rotation with respect to the distant stars.
A search is therefore made for variations of the
form 
\begin{equation}
A(\hat{t}) =  A_0 - A_1\sin(\Omega \hat{t} + \phi), 
\label{eq:4}
\end{equation}
where $A_0$, $A_1$ and $\phi$ are constants and are
extracted by measuring the asymmetry $A$ in Eq.~\eqref{raw} in bins of the
sidereal phase $\Omega\hat{t}$, and fitting to the value in each bin
with Eq.~\eqref{eq:4}. Measurements of $A_0$ and $A_1$ are then
interpreted as limits on $\Delta a_{\mu}$ (transverse 
$\Delta a_{\perp} = \sqrt{\Delta a_X^2+\Delta
a_Y^2}$, longitudinal $\Delta a_Z$, and time component $a_T$) from $\Bs$-$\barBs$
oscillations. A nonzero value of $\Delta a_Z$ and $\Delta a_T$ would
lead to a  CPTV asymmetry not varying with sidereal time.
 
\begin{figure}
\begin{center}
\includegraphics[width=\hsize]{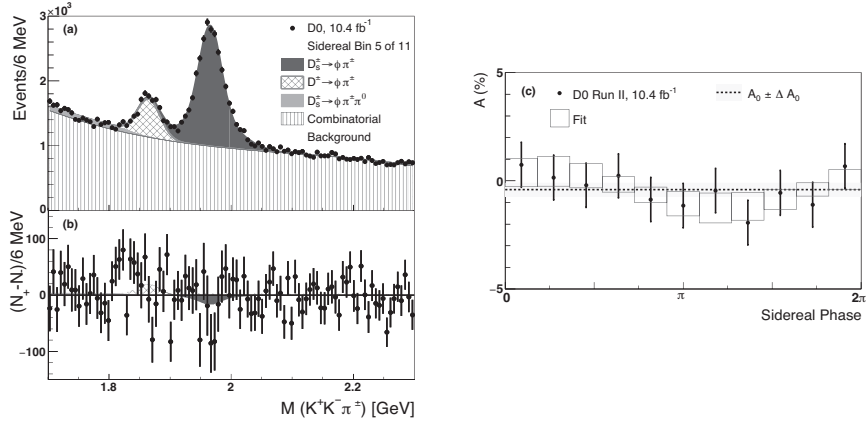}
\end{center}
\caption{(a) Reconstructed total signal and fit for yield $(N_+ + N_-)$ in one sidereal bin, (b) distribution of 
$(N_+ - N_-)$ and fit for the same sidereal bin, and (c) measured asymmetries $A(i)$ as a function of sidereal phase plus fit to test for variation with sidereal phase. }
\label{fig1}
\end{figure}

A typical fit to find the sum $(N_+ + N_-)$ and difference $(N_+ - N_-)$ yields of
$\Bsdecay X$ in a particular sidereal phase bin are shown in Fig.\ \ref{fig1}.
Figure~\ref{fig1}(c) then shows a fit testing for a sidereal phase dependence, finding
$A_0 = (-0.40 \pm 0.31)\%$ and $A_1 = (0.87 \pm 0.45)\%$, both consistent with zero
and hence exhibiting no significant evidence of Lorentz or CPT violation. From these results, a 95\% upper limit of
$\Delta a_{\perp} < 1.2 \times 10^{-12}$~GeV and 
two-sided confidence
interval of $(-0.8  < \Delta a_T - 0.396 \Delta a_Z < 3.9) \times 10^{-13}$~GeV are
extracted.\cite{resultpub}

\section{Discussion}

These results represent the first direct search for CPT-violating effects exclusively in the $\Bs$-$\barBs$ oscillation system. 
For CPTV to explain the difference between the D\O\ like-sign asymmetry\cite{dimuon2013} and the SM requires that $\Delta a_T - 0.396 \Delta a_Z$  be of the order of $10^{-12}$, implying that CPT violation is unlikely to contribute a significant fraction of the observed dimuon charge 
asymmetry.\cite{vanKooten}
These limits constrain a linear combination of the Lorentz-violating coupling constants $a^q_{\mu}$ for the $b$ and $s$ valence quarks in the \Bs\ meson that are different from the linear combinations of valence quarks in the $B^0$, $D^0$, and $K^0$ 
mesons.\cite{datatables}
As presented at this conference, a subsequent publication\cite{LHCb} from the LHCb Collaboration has improved on the previous best limits presented here by an order of magnitude primarily due to the much larger boost of the \Bs\ mesons at LHCb.

\section*{Acknowledgments}
We thank A.\ Kosteleck\'y for valuable conversations 
and also acknowledge support 
from the Indiana University Center for Spacetime Symmetries.


\begin{thebibliography}{xx}

\bibitem{CPT2}
V.A.\ Kosteleck\'y and S.\ Samuel, 
Phys.\ Rev.\ D {\bf 39}, 683 (1989);
V.A.\ Kosteleck\'y and R.\ Potting,  
Nucl.\ Phys.\ B {\bf 359}, 545 (1991).

\bibitem{cpt_preprint}
V.A.\ Kosteleck\'y and R.\ Van Kooten,
Phys.\ Rev.\ D {\bf 82}, 101702 (2010),
and references therein.

\bibitem{SME}
D.\ Colladay and V.A.\ Kosteleck\'y,
Phys.\ Rev.\ D {\bf 55}, 6760 (1997);
Phys.\ Rev.\ D {\bf 58}, 116002 (1998);
V.A.\ Kosteleck\'y,
Phys.\ Rev.\ D {\bf 69}, 105009 (2004).

\bibitem{CPT1}
V.A.\ Kosteleck\'y and R.\ Potting, 
Phys.\ Rev.\ D {\bf 51}, 3923 (1995).

\bibitem{dimuon2013} 
V.M.\ Abazov \etal, 
Phys.\ Rev.\ D {\bf 89}, 012002 (2014).

\bibitem{vanKooten}
V.A.\ Kosteleck\'y and R.\ Van Kooten, 
Phys.\ Rev.\ D {\bf 82}, 101702(R) (2010).

\bibitem{assl}
V.M.\ Abazov \etal, 
Phys.\ Rev.\ Lett.\ {\bf 110}, 011801 (2013).

\bibitem{resultpub}
V.M. Abazov \etal, 
Phys.\ Rev.\ Lett.\ {\bf 115}, 161601 (2015).

\bibitem{hfag} 
Y.\ Amhis \etal, arXiv:1412.7515.

\bibitem{datatables}
{\it Data Tables for Lorentz and CPT Violation,}
V.A.\ Kosteleck\'y and N.\ Russell,
2016 edition,
arXiv:0801.0287v9.

\bibitem{LHCb}
R.\ Aaij \etal, 
Phys.\ Rev.\ Lett.\ {\bf 116}, 241601 (2016).

\end{thebibliography}
\end{document}